\documentclass[sigconf]{acmart}

\usepackage{float}
\usepackage{ifthen}
\usepackage{tabularx}

\newcommand{\topline}{\hrule height 0.4pt \vskip 4.6pt\relax}
\newcommand{\botline}{\vskip 4pt \hrule height 0.4pt \vskip 4.6pt\relax}

\AtBeginDocument{%
  \providecommand\BibTeX{{%
    \normalfont B\kern-0.5em{\scshape i\kern-0.25em b}\kern-0.8em\TeX}}}

\newcommand{\MVpm}{\mbox{MV$_{\!\! pm}$}}

\copyrightyear{2023}
\acmYear{2023}
\setcopyright{rightsretained}
\acmConference[TAS '23]{First International Symposium on Trustworthy
Autonomous Systems}{July 11--12, 2023}{Edinburgh, United Kingdom}
\acmBooktitle{First International Symposium on Trustworthy Autonomous
Systems (TAS '23), July 11--12, 2023, Edinburgh, United
Kingdom}\acmDOI{10.1145/3597512.3599698}
\acmISBN{979-8-4007-0734-6/23/07}

\begin{document}

\title{Anticipating Accidents through Reasoned Simulation}


\newboolean{showauthors}
\setboolean{showauthors}{true} 
\ifthenelse{\boolean{showauthors}}{
  \newcommand{\blindreview}[2]{#1}
}{\newcommand{\blindreview}[2]{#2}}

\blindreview{
    \author{Craig Innes}
    \affiliation{%
      \institution{School of Informatics, University of Edinburgh}
      \streetaddress{} 
      \city{Edinburgh}
      \state{Scotland}
      \country{UK}}
    
    \author{Andrew Ireland}
    \affiliation{%
      \institution{School of Mathematical and Computer Sciences, Heriot-Watt University}
      \streetaddress{1 Th{\o}rv{\"a}ld Circle}
      \city{Edinburgh}
      \state{Scotland}
      \country{UK}}
    
    \author{Yuhui Lin}
    \affiliation{%
     \institution{School of Mathematical and Computer Sciences, Heriot-Watt University}
      \streetaddress{1 Th{\o}rv{\"a}ld Circle}
      \city{Edinburgh}
      \state{Scotland}
      \country{UK}}
    
    \author{Subramanian Ramamoorthy}
    \affiliation{%
      \institution{School of Informatics, University of Edinburgh}
      \streetaddress{} 
      \city{Edinburgh}
      \state{Scotland}
      \country{UK}}
}{\author{Anonymous Authors}}

\blindreview{
    \renewcommand{\shortauthors}{Innes, Ireland, Lin and Ramamoorthy}
}{}
\begin{abstract}
A key goal of the \textit{System-Theoretic Process Analysis} (STPA) hazard analysis technique is the identification of loss scenarios – causal factors that could potentially lead to an accident. We propose an approach that aims to assist engineers in identifying potential loss scenarios that are associated with flawed assumptions about a system’s intended operational environment. Our approach combines aspects of STPA with formal modelling and simulation. Currently, we are at a proof-of-concept stage and illustrate the approach using a case study based upon a simple car door locking system. In terms of the formal modelling, we use \textit{Extended Logic Programming}~(ELP) and on the simulation side, we use the CARLA simulator for autonomous driving. We make use of the problem frames approach to requirements engineering to bridge between the informal aspects of STPA and our formal modelling. 
\end{abstract}




\maketitle

\section{Introduction}\label{sect:intro}
\textit{System-Theoretic Process Analysis} (STPA) \cite{leveson2016engineering} extends traditional hazard analysis techniques to include a range of causal factors that can affect system safety. A key goal of STPA is the identification of causal factors that give rise to unsafe control actions, which in turn could lead to hazards and ultimately an accident –- what Nancy Leveson calls a \textit{loss scenario}. 

While STPA is systematic, it represents a relatively informal style of analysis. As a consequence, STPA 
provides no mechanized support to prevent errors such as inconsistencies and invalid assumptions entering into the
analysis. 
Given that loss scenarios often arise because of invalid environment assumptions \cite{leveson2016engineering},  
we believe that mechanized support for STPA is well motivated\footnote{ 
Such flaws can occur at design-time but can also arise as a consequence of changes to a system's operational environment.}.  

s
We propose a mechanized approach that uses formal modelling to verify the consistency between safety related constraints and assumptions. Such a verification relies upon the validity of the safety assumptions. To check validity we use simulation. Specifically, we search for environment conditions under which the safety assumptions are violated. Such violations represent loss scenarios. So while formal modelling ensures consistency of the analysis, we rely upon simulation to identify the concrete loss scenarios. 

Our proposed approach is at the proof-of-concept stage. We illustrate it using a case study based upon a simple car door locking system. To bridge between STPA and formal modelling we use the notion of a \textit{problem frame} \cite{jackson2001problem} combined with \textit{Extended Logic Programming} (ELP) \cite{gelfond1991classical}, while on the simulation side, we use the \textit{CARLA} (Car Learning to Act) 
simulator \cite{dosovitskiy2017carla} for autonomous driving. In section \ref{sect:bg}, we provide background on STPA and the aspects of ELP and problem frames that we make use of in this paper. We also provide an overview of the CARLA simulator and the features that we rely upon in this paper.

\section{Background}
\label{sect:bg}
We follow the 4-step description of STPA given in \cite{stpahandbook}.  The first step involves identifying the potential losses (L), system-level hazards (H) and system-level constraints (SC). A hazard denotes a system state coupled with environmental conditions that lead to a potential loss. The role of the system-level constraints is to guard against the identified hazards from occurring. Step 2 focuses on the achievement of the system-level constraints. First, it involves defining a control structure – a diagrammatic description of a system’s controllers and controlled processes, together with how they interact, i.e., control actions and feedback.  Second, the responsibilities for each controller are defined, where a responsibility (R) represents a refinement of a system-level constraint. Each responsibility is identified with part of a controller’s process model and control logic, which are informed by feedback from the processes that it controls. Finally, a control action is associated with each responsibility. Step 3 identifies how inadequate control could lead to a hazard by considering the following cases:
\begin{enumerate}
    \item A control action that when not applied causes a hazard.
	\item A control action that when applied causes a hazard.
	\item A control action that when applied too early, too late or out of sequence causes a hazard.
	\item A control action that is stopped too soon or applied too long causes a hazard.
\end{enumerate}
Note that the final case is applicable to non-discrete control actions. For each unsafe control action, a corresponding control constraint is defined, which specifies a property that must be satisfied in order to prevent the controller exhibiting the unsafe control actions. Figure 1 summarises the first 3-steps of STPA. Despite design-level verification, the fourth step focuses on identifying how unsafe control actions could occur and lead to the violation of system-level constraints. As noted above, these are referred to as loss scenarios. In \cite{leveson2016engineering}, the process of identifying a loss scenario is described as ‘the usual “magic” one that creates the contents of a fault tree.’ There are many ways in which unsafe control actions could occur. In section \ref{subsect:step4}, we focus on flawed assumptions as the source of loss scenarios.

\begin{figure*}[h]
    \topline
    \centering
    \includegraphics[width=0.8\textwidth]{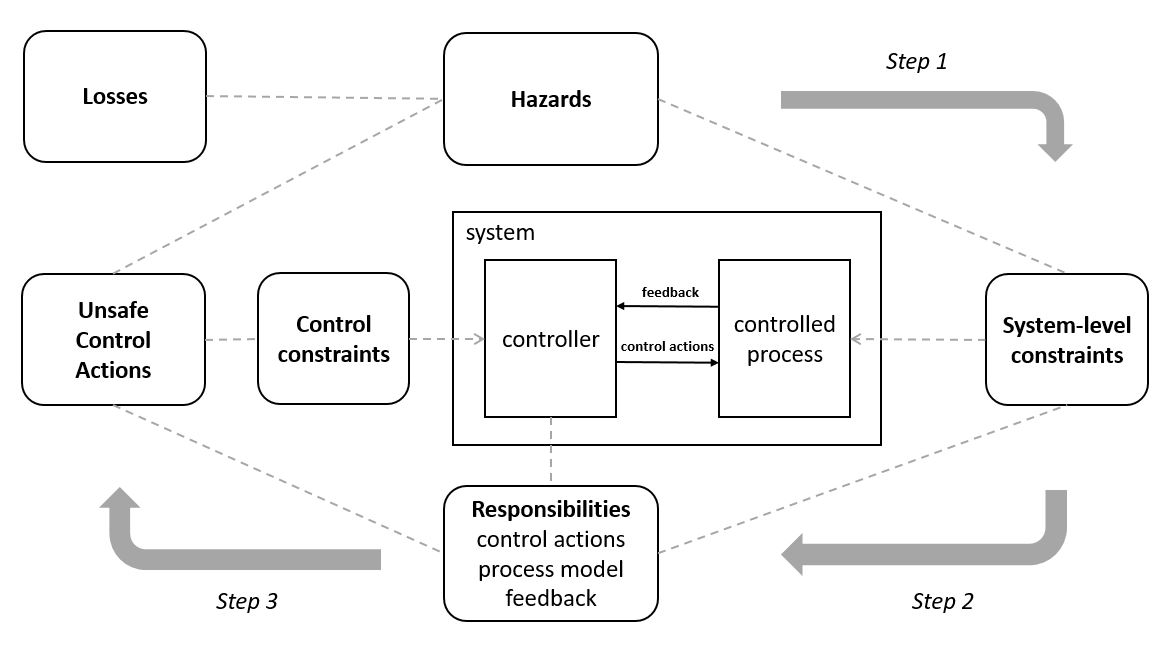}
    \begin{quote}
        The diagram above shows the first 3 steps of the STPA hazard analysis process -- from the definition of losses and hazards through to defining control actions and identifying how unsafe control actions can lead to hazards. Step 4, which is not shown in the diagram, focuses on the identification of loss scenarios (see section \ref{subsect:step4}).
    \end{quote}
    \botline
    \vspace*{-0.15in}
    \caption{Summary of steps 1-3 of STPA.}
    \label{fig:stpa}
\end{figure*}

As mentioned above, we use the notion of a problem frame \cite{jackson2001problem} to bridge between the informal and the formal. A problem frame provides a way of verifying the consistency between a system-wide requirement (i.e., system-level constraint) and the specification of a software component (i.e., control constraint). The problem frame representation has both graphical and logical elements. Substituting the STPA control structure with a problem frame involves formalizing the domain assumptions associated with the system and its operational environment.  An informal verification is represented by means of an informal argument diagram \cite{seater2007requirement} which will be illustrated in sections \ref{subsect:step3} and \ref{subsect:step4} 

In terms of formal logic, the problem frames approach is relatively agnostic. Our case study is quite simple, so we opt for classical propositional logic -- specifically we use \textit{Extended Logic Programming }(ELP) \cite{ gelfond1991classical}. As a consequence, we ignore delays between observations and actions. More expressive formalisms could be used. ELP represents an extension of logic programming, e.g., \textit{Prolog} \cite{giannesini1986prolog, clocksin2003programming}. Logic programming is a declarative programming paradigm that is based upon formal logic. A program is defined in terms of a collection of clauses, where a clause
is either a fact or a rule. This clausal form, which is a subset of first-order logic, takes the following
general form: 
\begin{equation}\label{eq: clause}
 L \verb+:-+ \, A_1, \; \ldots \; A_m, \; not \; B_1, \; \ldots \; not \;B_n
\end{equation}
where $A_i$ is a positive predicate and $B_j$ is a negated one. 
Program execution is based upon the resolution inference mechanism. 
From a reasoning perspective, logic programming is non-monotonic. That is, 
the fact that $\Gamma \vdash C$ holds does not guarantee that  $\Gamma, A \vdash C$ also holds. This is because 
negation (i.e., $not$) is evaluated by a non-monotonic rule called {\em negation-as-failure} (a form of the closed-world assumption). This means that if a statement cannot be derived then it is assumed to be false. ELP extends logic programming with a classical (or explicit) form of negation. While $not \; P$ denotes negation-as-failure, $-P$ denotes classical negation. To conclude $-P$ it is
not sufficient to have failed to derive $P$; explicit evidence for the negation of $P$ is required. 

Formal modelling is suited to relatively abstract models where the state space is small and can be exhaustively explored. In contrast, simulation allows one to selectively explore large state spaces. In terms of simulations, we have used CARLA \cite{dosovitskiy2017carla}, which is a simulation platform for testing the performance of automated vehicle systems. It provides a rich set of features for capturing the many complex interactions between the many integrated modules within a modern Autonomous Vehicle (AV) pipeline. In terms of physical dynamics and control, CARLA can simulate vehicles that are affected by a highly granular set of parameters. This can be as broad as overall mass, all the way down to gear ratio, drag coefficient of the chassis, tyre friction, and wheel stiffness. Also, CARLA provides libraries, e.g., Python API, to specify/program scenarios for both local control and global route planning of large numbers of vehicles on the road simultaneously.
A simulator provides a way to efficiently generate many thousands of example rollouts/scenarios of the AV system given a user-specified starting specification. So, while symbolic domain knowledge may provide heuristics about which configuration parameters are likely to be relevant to safety, testing variants of those parameters out in simulation allows us to see how they affect behaviour in practice, and thus search for falsifying examples.

\section{Related work}\label{sect:related}
As illustrated above, the problem frame approach provides a technique for verifying the correctness of a system-wide requirement with respect to a specification and domain assumptions. In \cite{seater2007requirement}  Alloy \cite{jackson2012software} is used to formalize and verify problem frame instances. Alloy uses first-order relational logic. Moreover, they use a technique called requirements progression to mechanically derive a specification from a given system-wide requirement and its associated domain assumptions. Here this would correspond to deriving control constraints from the system-level constraints and associated domain assumptions.  In section \ref{subsect:step4}, we use the notion of an anti-system-level constraint. This is similar to the notion of an anti-requirement introduced in \cite{lin2003introducing} to represent the intentions of a malicious user.

As a logic-based declarative language, ELP has been used as a formal modelling and problem-solving tool. ELP is supported by a knowledge representation and reasoning framework called Answer Set Programming (ASP) \cite{marek1999stable, gebser2012answer}. ASP has been applied to decision-making, planning, diagnostic reasoning and  configuration optimisation, e.g., decision support systems \cite{nogueira2001prolog}, configuration optimisation in railway safety systems \cite{aschinger2011optimization}, action planning in the setting of multiple robot collaboration tasks \cite{erdem2012answer}, and diagnosing failures of an automatic whitelisting system \cite{brik2015diagnosing}.

Combining formal reasoning and simulation is explored in \cite{gelman2014example}, where the SAL model checker is used to guide the use of agent-based simulation via WMC \cite{sierhuis2002modeling} Specifically, the SAL model checker was used to model a known automation surprise associated with the Airbus A320 autopilot, which was then subsequently explored via WMC. They found that WMC provided psychological plausibility to the counterexample generated by SAL. Moreover, WMC provided multiple refinements to the given SAL counterexample. With the growth in autonomous systems that rely upon human interaction, e.g., self-driving cars, then tools that assist in the early identification of automation surprises will have an important role to play within the assurance process. 
The use of formal methods in conjunction with STPA is not new. In \cite{howard2019methodology,colley2013formal} a methodology for incorporating the first three steps of STPA within the \textit{Event-B} \cite{abrial2010modeling} formal modelling tool is described. What distinguishes our proposal is the focus on identifying loss scenarios. Specifically, the use of formal modelling to identify abstract loss scenarios which are then elaborated via simulation.

Simulating autonomous vehicle behaviour requires coordinating multiple related, high-fidelity (computationally expensive) components. This includes sensor models (e.g., Cameras, Lidar, Radar), multibody vehicle dynamics, traffic flows, and pedestrian behavioural patterns \cite{yang2021survey}. In particular, such platforms distinguish between simulating the environment itself, and the environment as seen by the vehicle’s perception systems. Discrepancies between these views can be dangerous \cite{hoss2022review}. Such simulators constitute a core component in many verification tasks such as falsification and failure probability estimation \cite{corso2021survey}. AV Simulators allow us to efficiently replicate a given safety scenario with high fidelity, but it is impossible to test all possible road scenarios. Thus, the primary challenge is to identify which scenarios are worth simulating. Existing approaches to scenario selection involve either hand-crafted databases of safety-critical situations or data-driven scenario generators trained on large amounts of passively collected real-world traffic data \cite{stellet2020validation}.

\section{Adding Formal Modelling and Simulation to STPA}
\label{sect:fm and stpa}
To illustrate our proposal, we use an automatic car door locking example that involves a simple interlock mechanism. 
The example is inspired by Michael Jackson’s aircraft braking system example \cite{jacksonsoftware}, and is strongly related to an accident involving a fly-by-wire aircraft \cite{WarsawA320-93}. In our example, we are concerned with ensuring that while a car is moving its doors should be automatically locked and while the car is stationary, the doors should be automatically unlocked.
The task is to construct a Door Lock Control Unit (DLCU) that ensures the locking and unlocking of the doors occurs at the right time. A fundamental design assumption will be that when the car is moving the wheels will be turning. From the perspective of STPA, the controlled process involves the locking mechanism for the car's doors and the car's wheel sensors. The DLCU receives feedback in the form of wheel-pulse signals from the sensors when the wheels are turning. In terms of control, the DLCU will be able to send lockDoors and unlockDoors signals to the doors.
Below we follow the first three steps of the STPA hazard analysis technique as described above. We use the following propositions to bridge between the informal and formal: 

\begin{description}
\item [\rm{\textbf{WT}:}] true when car wheels are turning, otherwise false.
\item [\rm{\textbf{DL}:}]  true when car doors are locked, otherwise false.
\item [\rm{\textbf{WP}:}] true when there exists a wheel-pulse signal, otherwise false. 
\item [\rm{\textbf{DS}:}]   true when there exists a door-lock signal, otherwise false. 
\item [\rm{\textbf{MV}:}] true when the car is moving, otherwise false.
\item [\rm{\textbf{MV$_{pm}$}:}] true when the process model indicates the car is moving, otherwise false.
\end{description}
More realistically, these definitions would be developed hand-in-hand with the application of STPA.

\subsection{Define purpose and scope of the analysis (step 1)}\label{subsect:step1}
From the perspective of people travelling in a car, clearly, the loss of their lives or injuries due to safety failings would be personally catastrophic/devastating. In addition, the manufacturer would experience loss through the legal consequences of such safety failings, as well as potentially suffering reputational loss. For the purposes of the example, we will focus solely on the human perspective, i.e.,

\begin{tabular}{c l}
     \textbf{L-1}: & loss of life or injury.\\
\end{tabular}

\noindent We have associated the following two hazards to \textbf{L-1}: 

\begin{tabularx}{\linewidth}{c X}
\textbf{H-1}: & a passenger opens their door while the car is moving.\\
\textbf{H-2}: & a passenger is unable to open their door while the car is not moving.
\end{tabularx}

\noindent \textbf{H-1} may result in a person falling from a moving car leading to loss of life or injury, while \textbf{H-2} may prevent a person from leaving a stationary car when their safety is in jeopardy, e.g., the car is on fire. Following STPA, each hazard should be associated with a system-level constraint, i.e.

\begin{tabularx}{\linewidth}{c X}
\textbf{SC-1}:& if the car is moving then the doors must be locked $(MV \Rightarrow\ DL)$. [\textbf{H-1}] \\
\textbf{SC-2}:& if the car is not moving then the doors must be unlocked $(\neg MV \Rightarrow \neg DL)$. [\textbf{H-2}]
\end{tabularx}

\noindent Note that composing the above two system-level constraints gives:\\
\begin{tabularx}{\linewidth}{c X}
\textbf{SC-3}:& the car is moving if and only if the doors are locked  $(MV \Leftrightarrow\ DL)$. [\textbf{H-1, H-2}]\\
\end{tabularx}
\noindent Note that we will use \textbf{SC-3} for the purposes of verification and the identification of loss scenarios.   

\subsection{Model the control structure (step 2)}\label{subsect:step2}
\begin{figure*}[h]
    \topline
    \begin{center}
    \includegraphics[width=0.8\textwidth]{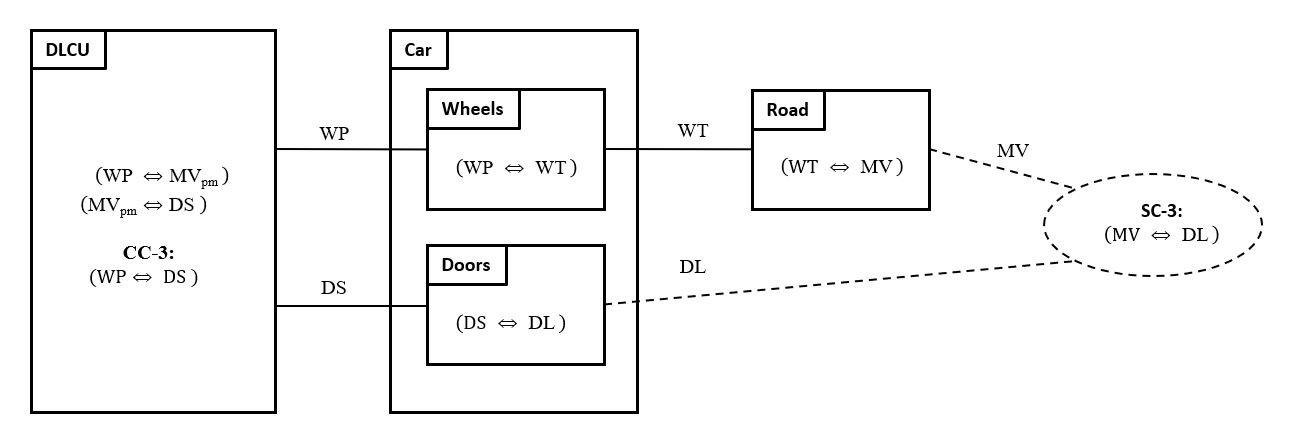}
    \end{center}
    \begin{quote}
        The above problem frame contains 3 domains, i.e., DLCU, Car and Road, where the Car domain has 2 subdomains, Wheels and Doors. The Road domain captures a key design assumption, i.e., $(WT \Leftrightarrow MV)$. Domains interact via shared phenomena, e.g., DLCU and Wheels share wheel-pulse signals (WP). The DLCU is a special domain; it represents the software controller, where its intended behaviour is specified via the control constraint CC-3. The dashed ellipse denotes the required system-level constraint, i.e., SC-3. The dashed lines indicate that the system-level constraint relates to the Road and Doors domains.
    \end{quote}
    \botline
    \vspace*{-0.15in}
    \caption{A Problem Frame for DLCU.}
    \label{fig:dlcu}
\end{figure*}

As noted in section 2, we use the notation of a problem frame rather than a control structure in order to bridge between the informal and the formal. Figure~\ref{fig:dlcu} provides a problem frame for DLCU. Note that the problem frame makes explicit the domain assumptions and shared phenomena that logically link the system-level constraint with the control constraint. Note in particular that the design assumption that the movement of the car is equivalent to its wheels turning is explicitly represented.

In terms of responsibilities, DLCU has two; each represents a refinement of a system-level constraint, i.e. 

\begin{tabular}{c l}
    \textbf{R-1}: &enable door locks when the car is moving. [\textbf{SC-1}]\\
    \textbf{R-2}: & disable door locks when the car is not moving. [\textbf{SC-2}]
\end{tabular}

As described in section \ref{sect:bg}, each responsibility is associated with elements of a controller's process model and control logic. These associations for DLCU are given below. They include both the STPA informal descriptions as well as a logical counterpart:

\begin{tabularx}{\linewidth}{l X}
\textbf{Responsibility}: &\textbf{R-1}\\
\textbf{Process model}:& the car is moving ($MV_{pm}$). \\
\textbf{Control logic}: &if wheel-pulse signal then process model indicates car is moving $(WP \Rightarrow\ \MVpm)$. \\ 
	& if process model indicates car is moving then lockDoors control action is applied $(\MVpm  \Rightarrow\ DS)$.\\
\textbf{Feedback}:& wheel-pulse signal (WP).\\
\end{tabularx}

\vspace{5mm}

\begin{tabularx}{\linewidth}{l X}
\textbf{Responsibility}: &\textbf{R-2}\\
\textbf{Process model}:& the car is not moving ($\neg \MVpm$). \\
\textbf{Control logic}: &if no wheel-pulse signal then process model indicates the car is not moving $(\neg WP \Rightarrow \neg \MVpm)$. \\
	& if process model indicates car is not moving then unlockDoors control action is applied \\  & $(\neg \MVpm \Rightarrow \neg DS)$. \\
\textbf{Feedback}:& no wheel-pulse signal $(\neg WP)$.
\end{tabularx}

The final part of step 2 involves defining the control actions that are intended to achieve a controller's responsibilities. In the case of DLCU there are two:

\begin{tabularx}{\linewidth}{c X}
     \textbf{CA-1}:& lockDoors (when door-lock signal then DS is true). $\text{[\textbf{R-1}]}$\\
     \textbf{CA-2}:& unlockDoors (when no door-lock signal $\neg DS$ is true). [\textbf{R-2}]
\end{tabularx}

\subsection{Identifying unsafe control actions (step 3)}\label{subsect:step3}
A STPA style hazard analysis for the two control actions associated with DLCU is shown in Table~\ref{tab:stpa-step3}. From the hazard analysis control constraints can be derived. The hazard analysis gives rise to the following two control constraints: \\
\begin{tabularx}{\linewidth}{c X}
    \textbf{CC-1}:& if wheel-pulse feedback then door-lock signal $(WP \Rightarrow\ DS)$. [\textbf{CA-1}]\\
    \textbf{CC-2}:& if no wheel-pulse feedback then no door-lock signal $(\neg WP \Rightarrow \neg DS)$. [\textbf{CA-2}]
\end{tabularx}

\begin{table*}[h]
    \centering
    \begin{tabular}{|c|c|c|c|c|} \hline
  \textbf{Control} & \textbf{Not applied}    & \textbf{Applied}         & \textbf{Applied too early,}      & \textbf{Stopped too soon} \\ 
\textbf{action}  & \textbf{causes a hazard} & \textbf{causes a hazard}  & \textbf{too late or out of}       & \textbf{or applied too long}  \\
                 &                          &                           & \textbf{sequence causes a hazard} & \textbf{causes a hazard}\\ \hline\hline
lockDoors   & Doors not locked        &  Doors locked when    & Doors locked too late   & \textit{N/A}  \\ 
   (\textbf{CA-1})     & when car moving  (\textbf{H1})   &  car stationary (\textbf{H2})  &      (\textbf{H1})                   &               \\ \hline
unlockDoors & Doors locked when       &  Doors unlocked       & Doors unlocked too soon & \textit{N/A}  \\
(\textbf{CA-2})  & car stationary    (\textbf{H2})  &  when car moving (\textbf{H1}) &      (\textbf{H1})                 &               \\ \hline  
        
    \end{tabular}
    \caption{Identifying Hazardous System Behaviour for DLCU.}
    \label{tab:stpa-step3}
\end{table*}

Note that composing the above two control constraints gives:
\begin{tabularx}{\linewidth}{c X}
\textbf{CC-3}:& wheel-pulse feedback if and only if door-lock signal $(WP \Leftrightarrow\ DS)$. [\textbf{CA-1, CA-2}]
\end{tabularx}

\begin{figure*}[h]
    \topline
    \centering
    \includegraphics[width=0.8\textwidth]{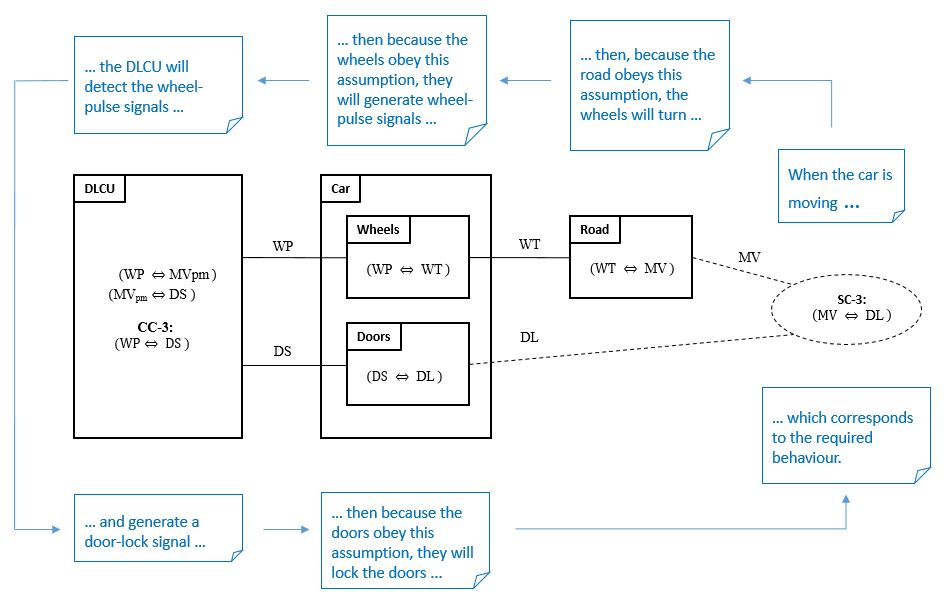}
    \includegraphics[width=0.5\textwidth]{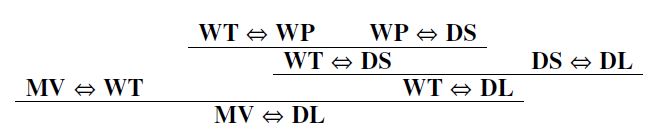}

\includegraphics[width=0.8\textwidth]{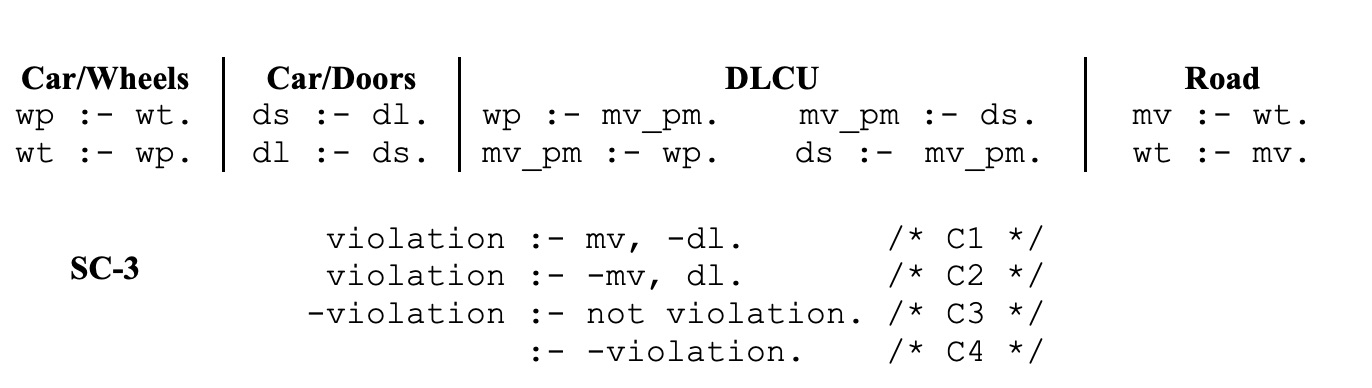}
\begin{quote}
Clauses C1 and C2 define how SC-3 can be violated. Note that the satisfaction of –dl and –mv requires explicit evidence for the negated propositions, i.e. failure to satisfy dl and mv is not sufficient to conclude that their negations hold. Clause C3 defines an explicit violation in terms of the failure satisfy violation. Clause C4 encodes the top-level proof: explicit evidence that there are no violation implies SC-3 it true. The Extended Logic Programming demos are available at \cite{res_elp}.
   \end{quote}
    \botline
    \vspace*{-0.15in}
    \caption{Proving consistency between CC-3 and SC-3.}
    \label{fig:proof-CC3-SC3}
\end{figure*}

The STPA Handbook emphasizes the need for verification, i.e. 
\begin{quote}
``STPA generates the safety requirements (e.g., \textbf{SC-3}) and constraints for the automated control algorithm (e.g., \textbf{CC-3}). These must, of course, be verified to be correct.''  
\end{quote}
As mentioned above in section \ref{sect:bg}, problem frames allow us to informally and formally verify the consistency between control constraints and system-level constraints. With regards to the consistency of \textbf{CC-3} and \textbf{SC-3}, Figure~\ref{fig:proof-CC3-SC3} provides an informal argument diagram, as well as a sequent style proof tree representation and our ELP representation.

\subsection{Identifying loss scenarios (step 4)}\label{subsect:step4}
As mentioned in section 2, we focus here on how loss scenarios could occur as a result of flawed assumptions. Specifically, we will focus on assumptions relating the environment in which a controller (i.e., DLCU) operates. Recall that a loss scenario results in the violation of a system-level constraint. In the case of our car example, the following proposition is equivalent to the violation of system-level constraint SC-3:
\begin{eqnarray}
\left(MV\ \land\ \lnot D L\right)\vee\left(\lnot M V\ \land\ DL\right) 
\label{form-1}
\end{eqnarray}
Note that the left disjunct of (\ref{form-1}) relates to \textbf{H-1}, a passenger opens their door while the car is moving, while the right disjunction relates to \textbf{H-2}, a passenger is unable to open their door while the car in stationary. Here we focus on the left disjunct, i.e.  
 \begin{eqnarray}
                    \left(MV\ \land\ \lnot D L\right)   
\label{form-2}
\end{eqnarray}           
The verification proof given in Figure 3 relies upon three assumptions. The violation of each assumption gives rise to a distinct loss scenario that satisfies (\ref{form-2}). Two of the scenarios correspond to component accidents, i.e., a wheel-pulse signal failure and a door-lock signal failure. Below we focus on the third assumption, i.e. 
 \begin{eqnarray}
        \left(WT\ \Leftrightarrow\ MV\right)
 \label{form-3}
 \end{eqnarray}
Specifically, we consider the case where (\ref{form-3}) is replaced by:  
\begin{eqnarray}
        \left(WT\ \Rightarrow\ MV\right)\ \land\ \lnot\left(MV\ \Rightarrow\ WT\right)   
\label{form-4}
\end{eqnarray}
 This violation of (\ref{form-3}) gives rise to a loss scenario where the car wheels are not turning but the car is moving, i.e.
\begin{eqnarray} 
    \left(\lnot W T\ \land\ MV\right)                                      
\label{form-5}
\end{eqnarray}

\begin{figure*}[h]
    \topline
    \centering
    \includegraphics[width=0.8\textwidth]{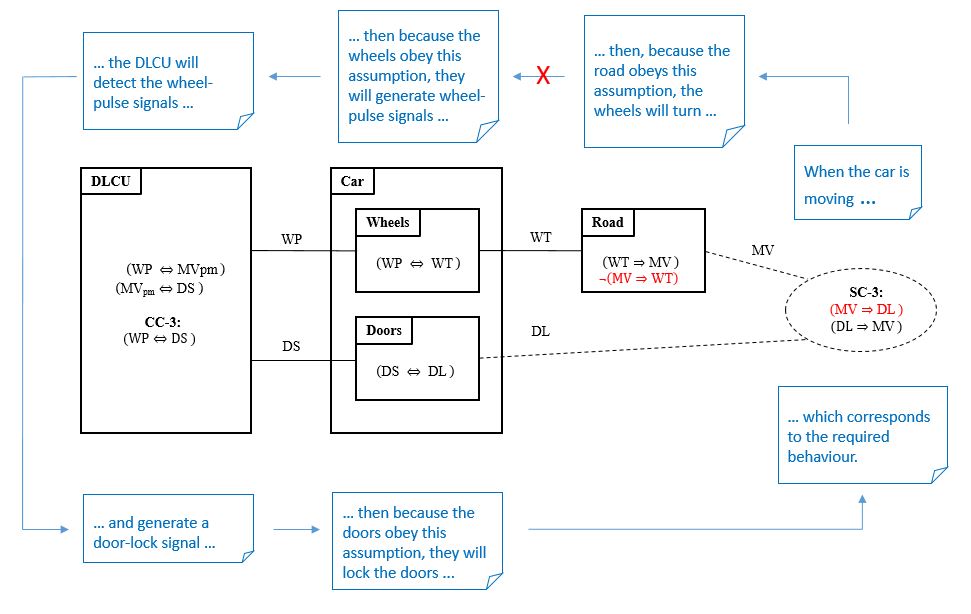}
    \includegraphics[width=0.6\textwidth]{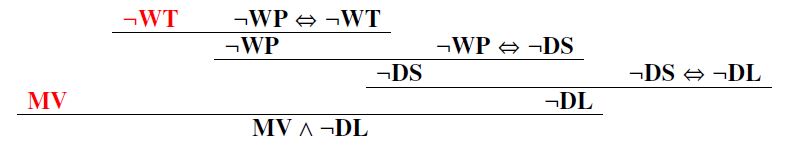}

    \includegraphics[width=0.8\textwidth]{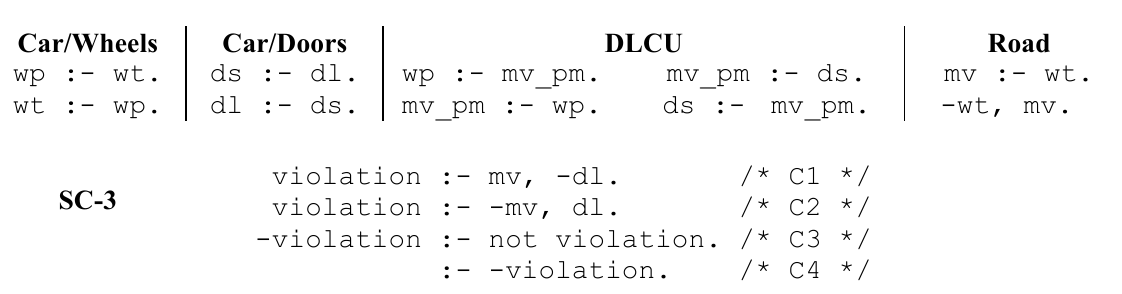}

\begin{quote}
The consequences of violating the design assumption, i.e., $\left(WT \Leftrightarrow MV\right)$,\ are shown above. Firstly, the informal argument diagram shows where the verification proof breaks. Secondly, the violation is represented as a sequent proof. Lastly, the ELP formulation of the violation is given.
    \end{quote}
    \botline
    \vspace*{-0.15in}
    \caption{A violation of the consistency between CC-3 and SC-3.}
    \label{fig:violation-CC3-SC3}
\end{figure*}

\begin{figure*}[h]
    \topline
    \centering
    \includegraphics[width=0.8\textwidth]{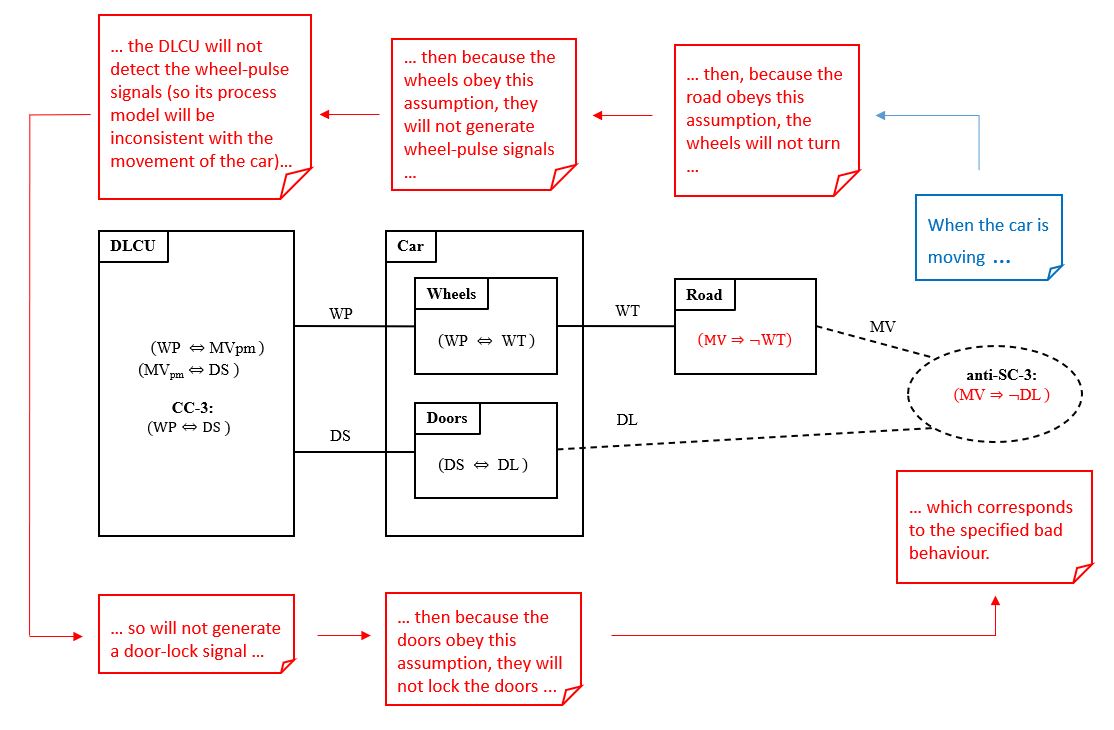}
    \includegraphics[width=0.6\textwidth]{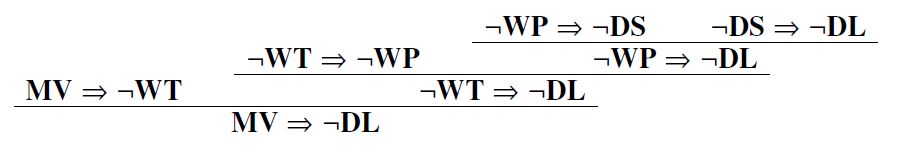}


    \includegraphics[width=0.8\textwidth]{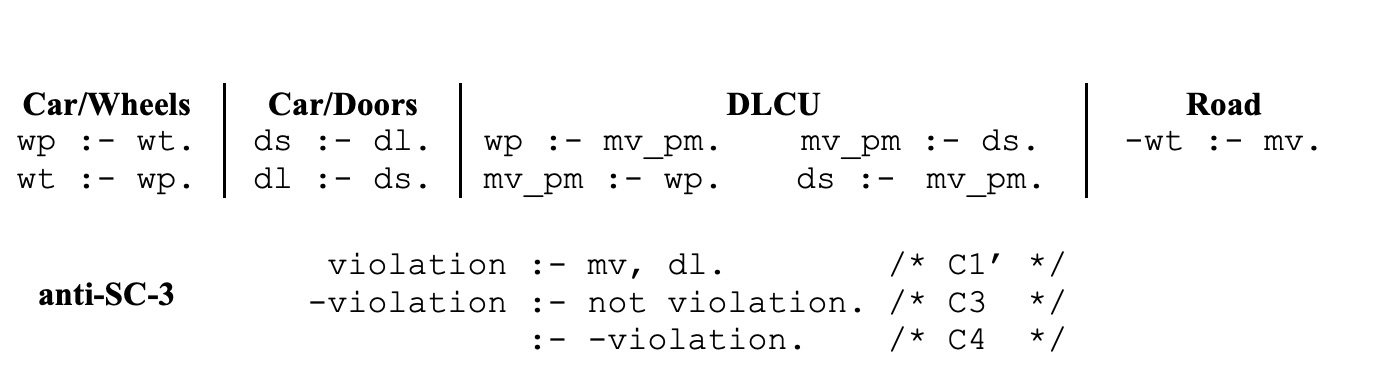}

\begin{quote}
Above the anti-system-level constraint, i.e., $(MV \Rightarrow \neg DL)$, is shown to be provable if the original design assumption, i.e., $(WT \Leftrightarrow MV)$, is replaced by $(MV \Rightarrow \neg WT)$. The first proof takes the form of an informal argument diagram. In addition, a sequent proof is provided along with the corresponding ELP formulation.
    \end{quote}
    \botline
    \vspace*{-0.15in}
    \caption{Proving consistency between CC-3 and anti-SC-3.}
    \label{fig:proving-CC3-anti-SC3}
\end{figure*}

As shown in Figure~\ref{fig:violation-CC3-SC3}, replacing assumption (\ref{form-3}) by (\ref{form-4}) (within the Road domain) breaks the verification proof given in Figure 3. In Figure~\ref{fig:proving-CC3-anti-SC3}, we present an alternative formal model of the loss scenario. We replace \textbf{SC-3} by an anti-system-level constraint, i.e., \textbf{anti-SC-3}. While a system-level constraint specifies, ‘something bad will never happen’, an anti-system-level constraint specifies, ‘something bad will happen’. The idea is that a scenario that achieves an anti-system-level constraint will denote a loss scenario. 

So formal modelling provides a mechanism for generating abstract loss scenarios. However, it provides no insight into the feasibility of the generated loss scenarios. We use simulation to provide such insight. Simulators are rich in terms of their knowledge about real-world phenomenon. What we propose is to leverage this knowledge in order to elaborate an abstract loss scenario. In so doing, we are exploiting synergies that exist between formal modelling and simulation, and providing engineers with more realistic and accessible loss scenarios to consider.
 
In our car example, (\ref{form-5}) provides the starting point for a simulation. Bridging the gap between the formal modelling 
(i.e., ELP) and simulation (i.e., CARLA) requires mapping the propositions $MV$ and $WT$ onto the corresponding parameters within the CARLA simulation space. Simply equating $MV$ with the car's velocity and WT with wheel speed is not sufficient. That is, it gives a 
low-fidelity approximation of the car's behaviour. CARLA provides various mechanisms to assist a developer to define a simulation. In terms simulation parameters, these are classified as follows:
\begin{description}
\item [Physics of a vehicle:] includes concepts such as torque, drag and friction.
\item [Traffic behaviour:] predefined behaviours are provided, i.e., cautious, normal, aggressive, but a developer is free to define the behaviour of the vehicles within their simulation.
\item [Perception:] weather conditions and the attributes of sensors, e.g., accuracy in poor lighting.
\end{description}
At the level of physical parameters, the following are relevant to our example:
\begin{itemize}
  \item Frictional coefficient of the road surface
  \item Velocity/acceleration of the vehicle in previous time step
  \item Tire friction
  \item Wheel torques
  \item Wheel speeds
  \item Engine RPM
  \item Drag coefficient on the vehicle body
  \item Normalised latitudinal and longitudinal forces at the wheels
\end{itemize}
To increase the fidelity of the approximation we explore the 8-dimensional factors listed above. 
Here, we combined the guidance provided by the formal model, i.e., $\neg WT \wedge MV$, and the relevant physical parameters from CARLA, to construct a potential loss scenario that violates (\ref{form-3}). The violation occurs when the velocity is greater than $0$ and the frictional coefficient of the road surface changes from $1.0$ to be less than or equal to $0.5$, i.e. the car is skidding.

Here the mapping was hand-crafted. To fully realize our proposal additional guidance is required in order to constrain CARLA. That is, we need access to domain knowledge about how stuff fits together as well as how stuff behaves in accordance with the laws of physics. For our illustrative example the concept of friction plays a key role in the loss scenario, i.e. wheel rotation requires friction between the car’s tyres and the road surface. Such knowledge could be represented in terms of an ontology. 

\begin{figure*}[h]
    \topline
    \centering
    \includegraphics[width=0.8\textwidth]{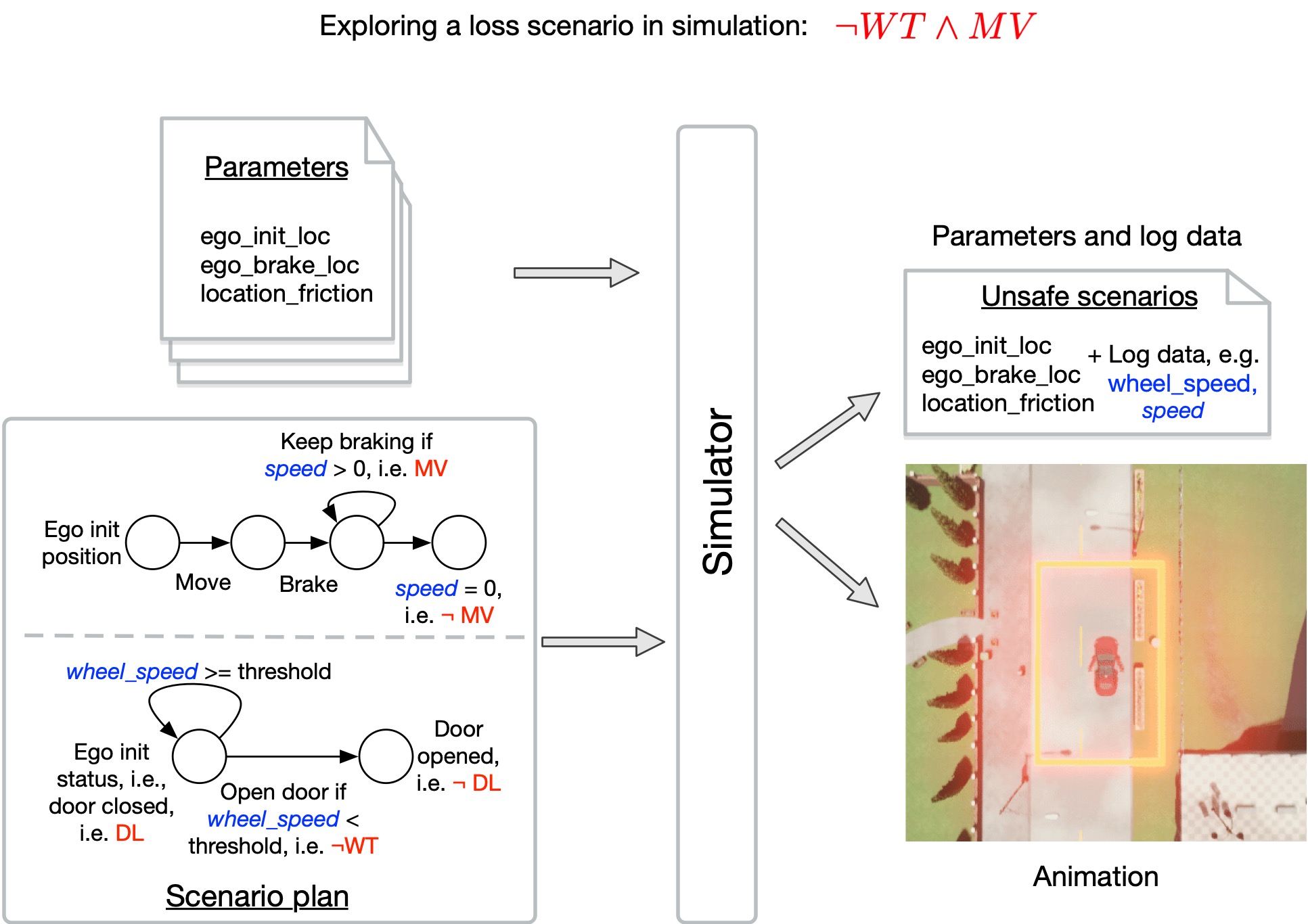}
    
\begin{quote}
 The inputs above on the left include, i) setup parameters for the simulation, and ii) template scenarios that encode the abstract loss scenario identified by the formal modelling. Note that while a graphical notation is shown, the developer uses a Python API to specify the inputs. The output from the simulator takes the form of i) an animation of the loss scenario, and ii) a log of all the relevant parameters. Note that a demo of the case study simulation is available via \cite{res_carla}, which includes the animations generated by CARLA.
    \end{quote}
    \botline
    \vspace*{-0.15in}
    \caption{Exploring loss scenarios using simulation.}
    \label{fig:loss-scenario}
\end{figure*}

\begin{figure*}[h]
    \centering
    \includegraphics[width=1.0\textwidth]{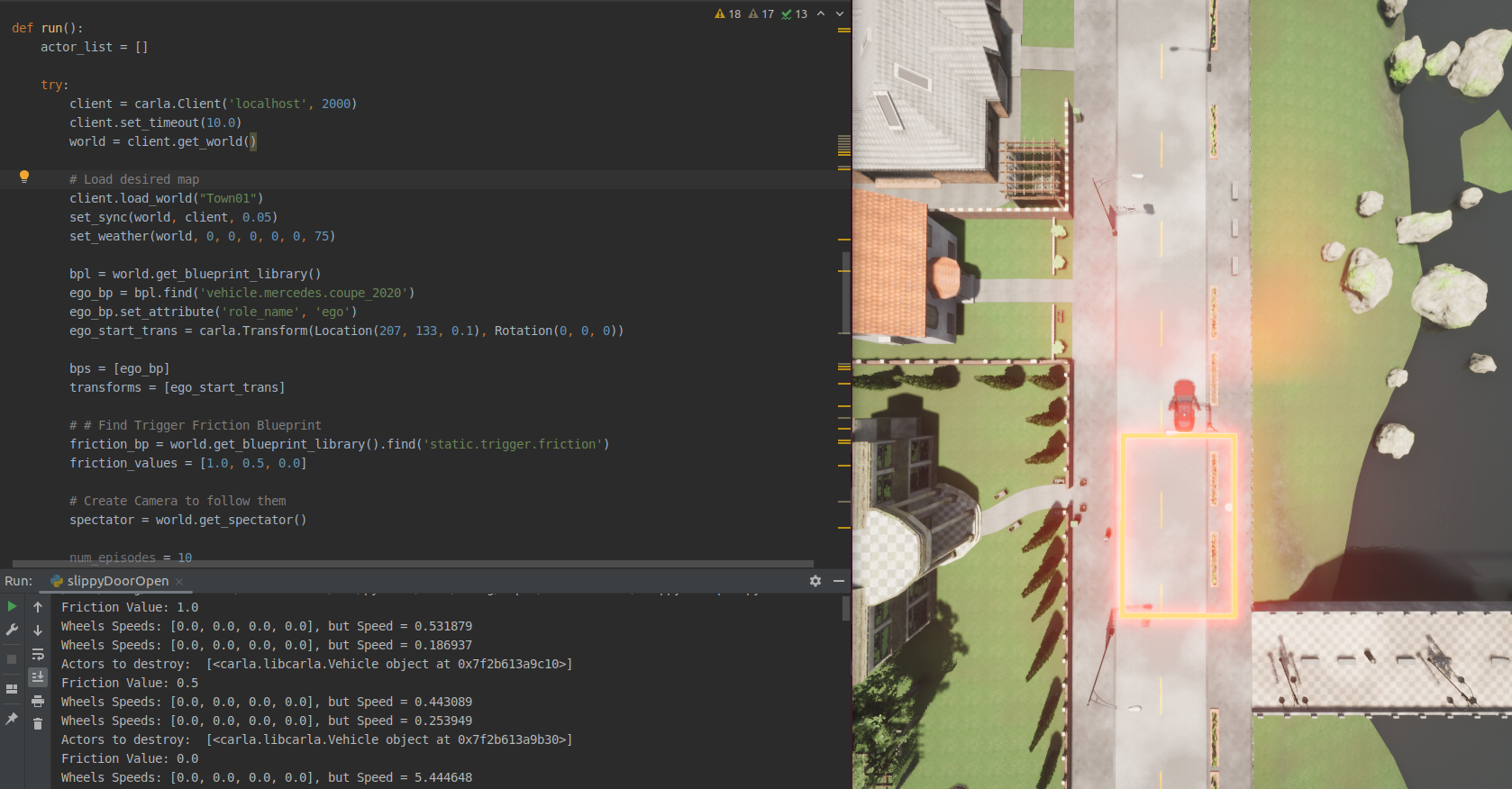}
    \vspace*{-0.1in} 
    \caption{A user’s view of the CARLA simulator.}
    \label{fig:CARLA-user-view}
\end{figure*}

From the perspective of the case study, our use of CARLA is summarised in Figure~\ref{fig:loss-scenario}. Note that the simulation demo and setup is available via \cite{res_carla}, which includes animations generated by CARLA. 
In Figure~\ref{fig:CARLA-user-view}, a snapshot of the CARLA simulation development environment is provided.

\section{Future work} \label{sect:future}
Two significant gaps need to be addressed in our future work. In order for our proposed approach to be effective and accessible, assistance is needed in bridging between STPA and ELP as well as ELP and CARLA. Tool support will be essential in managing the details. Full automation is not feasible or desirable. We envisage a level of dialogue between the tool and the user in order to ensure that each step in the translation is valid. In general, we believe that Interactive Task Learning \cite{laird2017interactive,appelgren2020interactive}  provides a promising approach to bridging these gaps. We are exploring mechanical analysis and vehicle models to develop hierarchical abstractions in order to bridge the gap between formal modelling statements and simulation variables, e.g., car moving ($MV$) and friction. As a consequence, we envisage the need for more expressive logics. Conversely, a less expressive simulation framework, e.g., CommonRoad \cite{CommonRoad17}, may serve our needs without the computational overheads of a simulator like CARLA. 
Finally, in addition to tool support, we also need to explore our ideas using more realistic case studies. 

\section{Responsible Research and Innovation}
\label{sec:rri}
Where innovation has safety related concerns, gaining public acceptance is an important challenge
that needs to be addressed. This is particularly true in the case of autonomous systems, such as self-driving cars. 
We believe that building public confidence in the way in which a system is designed will make a 
valuable contribution to addressing this challenge. Safety related defects are often introduced 
because of gaps between the traditional stages of the development process \cite{MarsPolar2000}. 
Our work seeks to reduce these gaps and therefore reduce the risk of such defects occurring. 
This argument requires a dialogue with the public, where evidence in the form of case studies 
will play a crucial role.  

\section{Conclusion}\label{sect:concl}
STPA is a popular technique that aims to assist engineers in analysing a system's control actions in relation to system-level hazards. The technique is structured and systematic but relatively informal compared to a formal method. Problem frames, if formalized, provides a mechanism for strengthening certain key informal aspects of STPA, i.e., verifying the consistency between safety related constraints and assumptions. Moreover, given such a formal verification we suggest that it can be used to identify loss scenarios related to invalid assumptions. Specifically, we propose the use of simulation to search for environment conditions under 
which safety related assumptions are violated. Engineers must be responsible for any safety related decisions, so 
ultimately it must be an engineer that judges whether or not a given loss scenario is feasible.
What we are proposing is an approach that assists engineers by providing them with simulated scenarios 
to review that are realistic and accessible.

\begin{acks}
The research reported in this paper is supported by the UKRI Trustworthy Autonomous Systems Node in Governance and Regulation (EPSRC grant EP/V026607/1).
\end{acks}

\bibliographystyle{ACM-Reference-Format}
\bibliography{ssc}

\end{document}